\documentclass[12pt]{article}
\title{Phase-Aligned Spectral Filtering for Decomposing Spatiotemporal Dynamics }
\author{Lu Meng \and Tian Zheng}
\date{Department of Statistics \\ Columbia University\\ \vspace{0.5cm} \today}

\usepackage{setspace} 
\usepackage{amsmath}
\usepackage{bbm}
\usepackage{amsfonts}
\usepackage{amssymb}
\usepackage{graphicx}
\usepackage{url}
\usepackage{hyperref}
\usepackage[linesnumbered,ruled,vlined]{algorithm2e}
\usepackage{enumerate}
\usepackage{longtable}
\usepackage{named}
\usepackage{dsfont}
\usepackage{subfig}

\renewcommand{\vector}[1]{\boldsymbol{#1}}
\newcommand{\E}{\mathbb{E}}
\newcommand{\R}{\mathbb{R}}

\newcommand{\I}{\mathds{1}}

\newcommand{\fil}[1]{\tilde{#1}}

\newcommand{\C}{\mathbf{C}}
\newcommand{\B}{\mathbf{B}}

\newcommand{\Z}{\boldsymbol{Z}}
\newcommand{\X}{\boldsymbol{X}}
\newcommand{\z}{\boldsymbol{z}}
\newcommand{\x}{\boldsymbol{x}}
\newcommand{\D}{\mathcal D_{m}}
\newcommand{\T}{\mathcal T_{n}}

\newcommand{\Arg}{\mbox{Arg}}

\begin{document}

\maketitle

\begin{abstract}
Spatiotemporal dynamics is central to a wide range of applications from climatology, computer vision to neural sciences. From temporal observations taken on a high-dimensional vector of spatial locations, we seek to derive knowledge about such dynamics via data assimilation and modeling. It is assumed that the observed spatiotemporal data represent superimposed lower-rank smooth oscillations and movements from a generative dynamic system, mixed with higher-rank random noises. Separating the signals from noises is essential for us to visualize, model and understand these lower-rank dynamic systems. It is also often the case that such a lower-rank dynamic system have multiple independent components, corresponding to different trends or functionalities of the system under study. In this paper, we present a novel filtering framework for identifying lower-rank dynamics and its components embedded in a high dimensional spatiotemporal system. It is based on an approach of structural decomposition and phase-aligned construction in the frequency domain. In both our simulated examples and real data applications, we illustrate that the proposed method is able to separate and identify meaningful lower-rank movements, while existing methods fail.
	
\end{abstract}

\section{Introduction}
The assimilation of spatiotemporal data is critical to the scientific discovery in a wide range of fields such as environmental sciences where temporal data are collected by spatially distributed remote-sensing platforms and sensor networks, and neural sciences where time series of brain activities are measured using images from functional Magnetic Resonance Imaging (fMRI) or electroencephalography (EEG) signals. Most such analyses are descriptive. In other words, they employ statistical models that would shed lights on the spatially dependent evolving processes of interests. Currently, there are two main approaches of spatiotemporal modeling (see \cite{wikle2010general} for an overview): i) via joint space-time covariance functions
\cite{cressie1999classes,gneiting2002nonseparable,ma2003families,wikle2003hierarchical,stein2005space,paciorek2006spatial,fonseca2011general,bevilacqua2012estimating,hsu2012group,choi2013nonparametric}; 
 ii) via direct dynamic models that uses either time-varying spatial models or spatially structured multivariate time series models
\cite{solna1996time,wikle1999dimension,huang2004modeling,xu2005kernel,gelfand2005spatial,johannesson2007dynamic,sigrist2012dynamic,gladish2014physically}.
The latter is often preferred in practice since it represents, comparing to the former, a more direct systems-oriented approach in connection with the scientific context concerning the spatiotemporal processes of interest. It can also be more flexibly integrated with stochastic methods and dynamic programming algorithms.  

Current technologies have enabled faster and denser data collection in both time and space. A major challenge in the analysis of today's spatiotemporal data is their high dimensionality. On the other hand, it is believed that the true spatiotemporal system being measured is smooth in space and time, rendering a lower-rank underpinning dependence structure for the high-dimensional observations. Identification of this lower-rank structure will therefore lead to scientific insights. Furthermore, in applications that involve predictive modeling using high-dimensional spatiotemporal data, information-preserving dimension reduction is of utmost importance to the construction of a reliable predictor. 

Existing dimension reduction methods for independent observations largely fall into two categories. {\em variable selection} methods (e.g., via regularization as in LASSO \cite{tibshirani1996regression}) that reduce the dimension of the original variable space by selecting a subset of the most ``important'' variables. {\em Filtering} or {\em transformation} methods (e.g. factor analysis~\cite{thurstone1931multiple}, Kalman filter \cite{kalman1960new}, independent component analysis~\cite{comon1994independent} and {\em etc}.), on the other hand, identify a low-dimensional manifold (i.e., a transformed feature space) that carries a substantial amount of the original information. Principal component analysis (PCA) \cite{pearson1901liii} is one of the most popular tools for dimension reduction via transformation (see \cite{jolliffe2002principal} for examples). It is also known as Hotelling transform \cite{hotelling1933analysis}, discrete Karhunen-Lo\`eve transform (KLT) \cite{karhunen1947lineare,loeve1948fonctions}, and empirical orthogonal functions (EOF) \cite{lorenz1956empirical}. PCA explores the covariance structure among the elements of a multivariate random vector and performs linear transformations such that the transformed variables, which are called principal components, are linearly uncorrelated and carry a maximal amount of original information. In most applications, a small number of leading components with the highest variances preserve a large portion of the overall variability in the original multivariate random vector. 

For temporal observations on a high-dimensional vector of spatial locations, however, the aforementioned methods for independent observations fail to account for temporal dependence that is possibly coupled with spatial dependence. For example, by treating observations at different time points as independent, principal components can be derived from spatiotemporal data. These principal components are contemporaneously uncorrelated but will exhibit autocorrelation and cross-autocorrelation at different time lags, which are hard to interpret and model. As a result, one can find a long list of filtering methods in the literature that are specifically designed for multivariate time series analysis \cite{box1977canonical,molenaar1985dynamic,back1997first,forni2000generalized,stock2002forecasting,matteson2011dynamic,wang2013sparse,chang2014segmenting,forni2015dynamic}. In particular, the generalized dynamic factor model proposed by \cite{forni2000generalized} decomposes the original multivariate process into moving averages of orthonormal white noises of a lower dimension. However, the model was not intended to provide a structural interpretation \cite{forni2000generalized}.
The singular spectral analysis by \cite{ghil2002advanced} decomposes the higher-dimensional series into additive components presenting trend, seasonal component and noise respectively. 
Matteson and Tsay (2011) \cite{matteson2011dynamic} directly looked for a contemporaneous linear transformation so that the transformed variables, which they call dynamic orthogonal principal components, have no linear and quadratic cross-correlation over time.
Chang {\em et al.\ } (2014) \cite{chang2014segmenting} also searched for a contemporaneous linear transformation so that the transformed multivariate time series form a group structure where any pair of transformed series from different groups will exhibit no cross-correlation. 
Neither~\cite{matteson2011dynamic} or~\cite{chang2014segmenting} considered possible lags between original variables and the latent factors. Furthermore, the targeted transformations are not guaranteed to exist. None of these methods can discover interpretable spatially evolving dynamic components, as we will show using extensive simulations. 

\begin{figure}[h]
	\centering
	\includegraphics[width=0.85\textwidth]{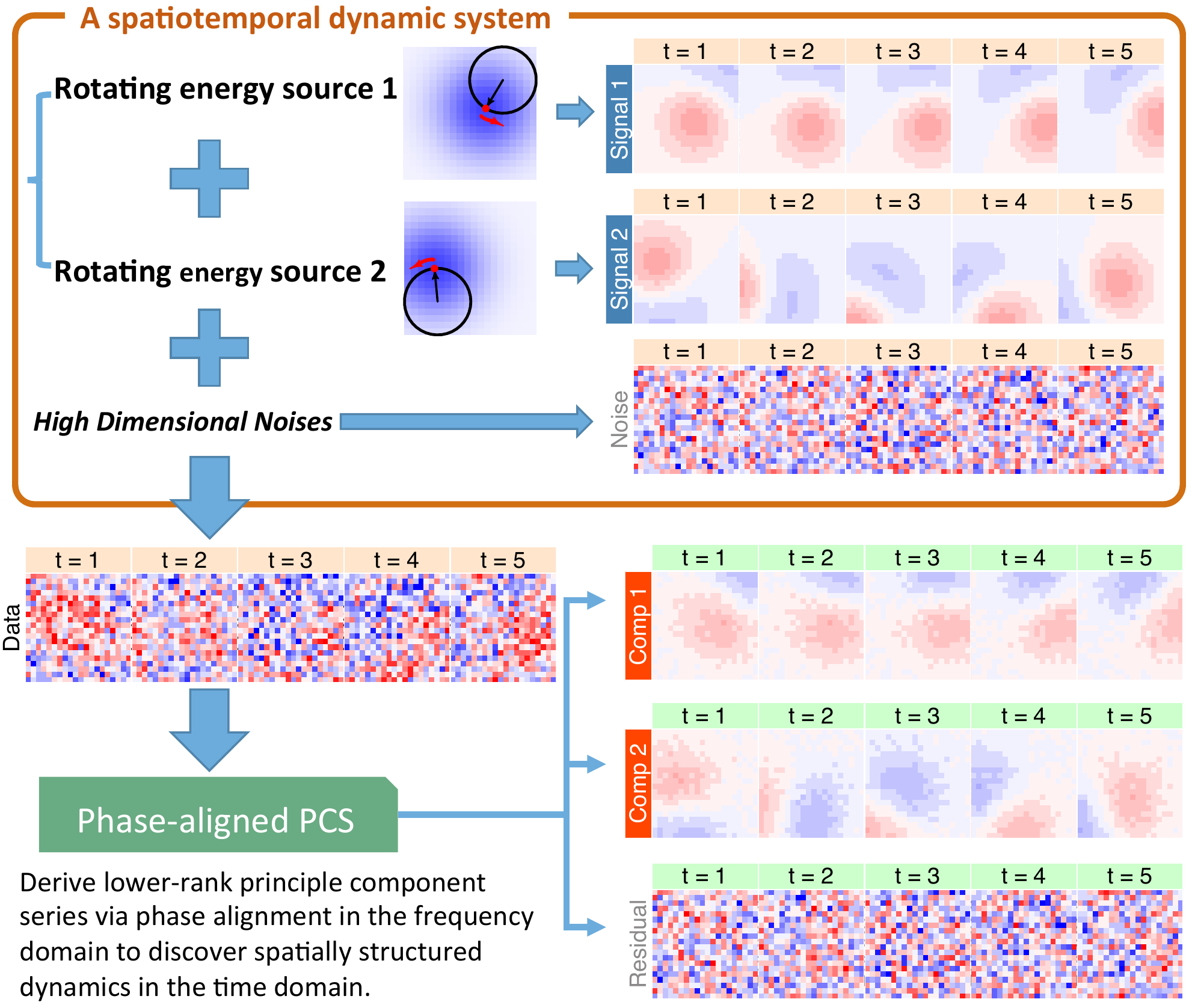}
	\caption{The proposed phase-aligned spectral filtering method discovers individual dynamic components from spatiotemporal data that consist of lower-rank smooth dynamic signals and high-dimensional noises.}
	\label{fig:diagram}
\end{figure}

To address the dimension reduction and signal decomposition problem for multivariate time series, \cite{brillinger1981time} proposes a spectral approach that directly decompose the spectral density matrices in the frequency domain and construct two sets of linear filters based on selected eigenvectors. The first set of filters transform the original high-dimensional series to low-dimensional \emph{principal component series} with zero coherence among each other. The second set transforms the principal component series back to a lower-rank component of the original data.  It leads to the most efficient dimension reduction for multivariate time series in terms of minimum mean squared error. Despite its elegant mathematical rigor, it has not been widely used in practice due to the lack of interpretability of the filtered principal component series and components. In particular, it is not clear how one should select eigenvectors at different frequencies to assemble interpretable filtered components. In terms of spatiotemporal data, for example, an interpretable component should be an oscillating dynamics in the space, which could correspond to different energy level and loadings (eigenvectors) at different frequencies. Therefore, using the eigenvector corresponding to the $k$-th largest eigenvalue at every frequency, as suggested in \cite{brillinger1981time}, to assemble filters is not well motivated. 

In this paper, we show that a number of spatial dynamic systems such as {\em phase propagation} result in a spatially-structured signature in the {\em complex argument} (or {\em phase offset}) of their {\em Fourier transform}, which is preserved across frequencies up to a linear transformation. Taking advantage of this result, we propose a spatial phase-aligned algorithm for constructing phase-aligned spectral filters, adapting the spectral approach of \cite{brillinger1981time} to spatiotemporal data (Figure~\ref{fig:diagram}). Eigenvectors from different frequencies in the {\em frequency domain} are clustered based on their complex argument to create filters that deliver interpretable spatial dynamics in the {\em time domain}. In both simulations and empirical applications, the proposed phase-aligned spectral filtering method returns clean and interpretable lower-rank spatiotemporal dynamics that explain a substantial proportion of the observed data. 

\section{Methodology}
\label{sec:method}
\subsection{Phase-aligned spectral filtering: an outline.}
\begin{figure}
	\centering
	\includegraphics[width=\textwidth]{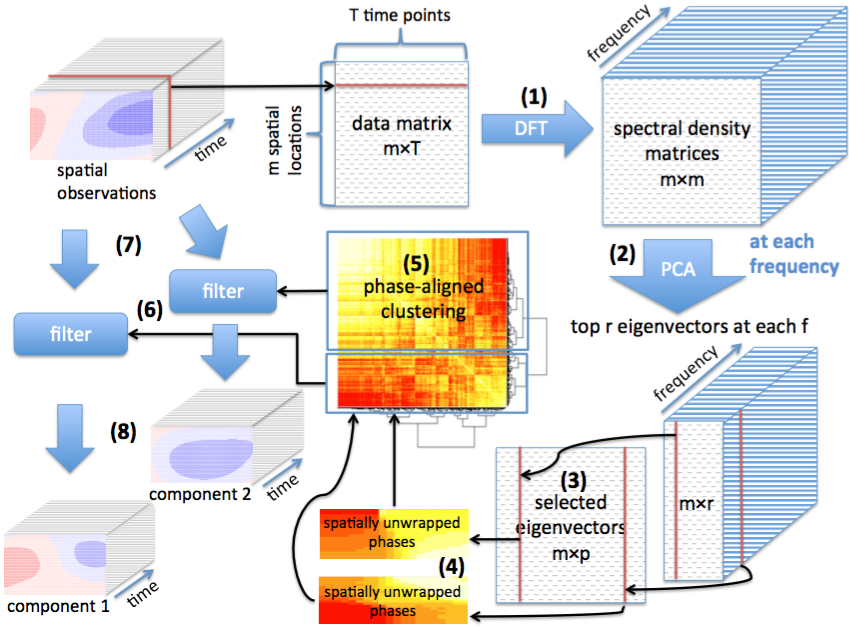}
	\caption{Outline of the phase-aligned spectral filtering decomposition of spatiotemporal dynamics. From vectorized spatial observations over time, we first apply {\bf (1)} discrete Fourier transform (DFT) to derive spectral density matrices. Applying {\bf (2)} principal component analysis at each frequency, we select eigenvectors corresponding to the top $r$ eigenvalues. We further impose {\bf (3)} shrinkage on the selected eigenvectors to remove those corresponding to eigenvalues below a predetermined threshold. Raw phase-values of selected eigenvectors are then {\bf (4)} unwrapped spatially and then {\bf (5)} clustered. A pair of linear filters based on eigenvectors in a cluster were {\bf (6)} constructed. Filtering the originally observed spatiotemporal data via two consecutive steps {\bf (7)} and {\bf (8)}, we derive separated low-rank spatially smooth dynamic components of the observed system. More details on these steps can be found in Section~\ref{sec:pasf}. }
	\label{fig:algorithm}
\end{figure}  
Our proposed {\em phase-aligned spectral filtering} method have two main motivations. The first motivation is Brillinger's dimension reduction strategy for multivariate time series using eigenvalue decomposition applied on the spectral density matrices in the frequency domain \cite{brillinger1981time}. By assembling linear filters based on eigenvectors from different frequencies, we can derive principal component series (PCS) that have zero coherence at all frequencies. This is a desirable properties for further predictive modeling using such series. To maintain the maximal information in the original series while reducing dimensions, eigenvectors corresponding to top eigenvalues are selected from each frequency. Applying a second set of filters constructed using the conjugate of of these selected eigenvectors at different frequencies, the principal component series can be filtered back into lower-rank components of the original data. In \cite{brillinger1981time}, it is suggested that the $k$th eigenvector from each frequency be used to assemble the filters for the $k$th component. However, this is not well motivated nor required for having zero coherence among the PCS or the rest of Brillinger's results in \cite{brillinger1981time}. As our second motivation, in this paper, we identify a link between phase-correlation in spectral densities and a number of low-rank smooth spatiotemporal dynamics.  This provides a novel approach for assembling spectral filters that will decompose the original high-dimensional time series into lower-rank spatially smooth dynamic components. The proposed framework follow a sequence of steps as shown in Figure~\ref{fig:algorithm}. 

In the following sections, we explain in details the phase-aligned spectral flitering method, starting with dimension reduction in the frequency domain (Section~\ref{sec:dr}), eigenvalue shrinkage (Section~\ref{sec:fdr}) and reassembling of principle component series (Section~\ref{sec:rs}). Section~\ref{sec:pasf} explains the computational steps of decomposing a high-dimensional spatiotemporal data set using the proposed phase-aligned spectral filtering method. 

\subsection{Dimension reduction in the frequency domain}\label{sec:dr}
Consider a real-valued spatiotemporal process 
$\{Z_{s,t}\in \R: s\in\D, t\in\mathcal T\}$ 
where the discrete spatial domain 
$\mathcal D_m = \{s_1,...,s_m\}\subset \mathcal D$ 
contains a set of $m$ locations and $\mathcal T=\{0, \pm 1, \pm 2, \cdots\}$.
Assume that $\Z_t = \{ Z_{s,t}: s \in \mathcal D_m\}$ is a second-order stationary $m$-dimensional vector-valued time series with mean $\vector \mu$ and $m\times m$ spectral density 
$f_{zz}(\omega)$ for $\omega \in [-\frac 1 2 , \frac  1 2]$. Without loss of generality, we assume $\vector \mu = \vector 0$. The spectral density matrix $f_{zz}(\omega)$ is a complex-valued positive semidefinite Hermitian matrix in the form of
\begin{equation}
f_{zz}(\omega) = \sum^{\infty}_{h=-\infty}\Gamma(h)e^{-2\pi i\omega h},
\end{equation}
where $\Gamma(h)$ is the autocovariance function of $\Z_t$.

Brillinger (1981)\cite{brillinger1981time} stated that for each integer $1\leq r\leq m$, there exist a pair of filters $\{\C^{[r]}_{\tau}\in\R^{r\times m}:u\in\mathcal T\}$ and  $\{\B^{[r]}_{\tau}\in\R^{m\times r}:u\in\mathcal T\}$ that minimizes the mean squared error 
\begin{equation} \label{eq:mini}
\E \overline{(\Z_{t}-\fil{\Z}_{t})^\top}(\Z_{t}-\fil{\Z}_{t})
\end{equation}
where 
\begin{equation}
\X^{[r]}_{t} = \sum^{\infty}_{\tau=-\infty} \C^{[r]}_{t-\tau} \Z_{\tau},
\mbox{ and, } 
\fil{\Z}_{t} = \sum^{\infty}_{\tau=-\infty} \B^{[r]}_{t-\tau} \X^{[r]}_{\tau}. 
\label{eq:brillinger}
\end{equation}
The sequence of matrices are
\begin{align}
\C^{[r]}_{\tau} &= \int^{1/2}_{-1/2} \C^{[r]}(\omega)\exp(2\pi i\tau\omega)d\omega, \label{eq:filter.c} \\
\B^{[r]}_{\tau} &= \int^{1/2}_{-1/2} \B^{[r]}(\omega)\exp(2\pi i\tau\omega)d\omega, \label{eq:filter.b}
\end{align}
where
\begin{equation*}
\C^{[r]}(\omega) 
\begin{pmatrix}
\overline{\vector v_{(1)}(\omega)^\top} \\
\vdots \\
\overline{\vector v_{(r)}(\omega)^\top}
\end{pmatrix}, \mbox{ }
\B^{[r]}(\omega) \left(\vector v_{(1)}(\omega), \cdots, \vector v_{(r)}(\omega)\right) = \overline{\C^{[r]}(\omega)^\top},
\end{equation*}
and $\vector v_{(1)}(\omega), \cdots, \vector v_{(r)}(\omega)$ are the $r$ eigenvectors of $f_{zz}(\omega)$ corresponding to its first $r$ largest eigenvalues $\lambda_{(1)}(\omega), \cdots, \lambda_{(r)}(\omega)$. The resulted $\X^{[r]}_{t}$, referred to as principal component series, has a spectral density matrix
\begin{equation}
f_{xx}(\omega) = 
\begin{pmatrix}
\lambda_{(1)}(\omega) & & \vector 0 \\
&\ddots& \\
\vector 0 & &\lambda_{(r)}(\omega)
\end{pmatrix},
\end{equation}
which means any pair of the principal component series $X_{k,t}$ and $X_{j,t}$ for $k\neq j$ has zero coherence.

\subsection{Further dimensionality reduction by shrinkage} \label{sec:fdr}
At each frequency, we decompose the spectral density as follows
\begin{align*}
	f_{zz} (\omega) =& \sum_{j=1}^{m}\lambda_{(j)}(\omega)\vector v_{(j)}(\omega) \overline{\vector v_{(j)}(\omega)^{\top}} \\
	=&
	\sum_{j=1}^{r}\lambda_{(j)}(\omega)\vector v_{(j)}(\omega) \overline{\vector v_{(j)}(\omega)^{\top}}\\
	&+ \sum_{j=r + 1}^{m}\lambda_{(j)}(\omega)\vector v_{(j)}(\omega) \overline{\vector v_{(j)}(\omega)^{\top}}\\
	=& f_{\fil{z}\fil{z}} (\omega) + f_{\epsilon\epsilon}(\omega).
\end{align*}
where $f_{\epsilon\epsilon}(\omega)$ is the spectral density of the error series $\vector \epsilon_{t} = \Z_{t} - \fil{\Z}_{t}$.
In many cases, the spectral power concentrates in a relatively small area along the frequency axis, that is, if we pool the first $r$ largest eigenvalues at all frequencies, there remain a large proportion of eigenvalues being relatively small, or even close to zero. Here we propose a threshold $\Delta$ and further define
$$
r_\omega = \max\{j: \lambda_{(j)}(\omega)\geq\Delta\}.
$$
The decomposition of the spectral density of $\Z_t$ becomes
\begin{align*}
	f_{zz} (\omega) 
	=&
	\sum_{j=1}^{r_\omega}\lambda_{(j)}(\omega)\vector v_{(j)}(\omega) \overline{\vector v_{(j)}(\omega)^{\top}}\\
	& + \sum_{j=r_\omega + 1}^{m}\lambda_{(j)}(\omega)\vector v_{(j)}(\omega) \overline{\vector v_{(j)}(\omega)^{\top}} \\
	&\overset{\Delta}{=} f_{\fil{z}^\star\fil{z}^\star} (\omega) + f_{\epsilon^\star\epsilon^\star}(\omega).
\end{align*}
If $r_\omega$ is 0, then $f_{\fil{z}^\star\fil{z}^\star} (\omega)=0$.
Compared to the previous filtered series $\fil{\Z}_t$, the new filtered series $\fil{\Z}^\star_t$ is less noisy and more robust. This shrinkage step is also important and necessary in the next phase-aligned reassembling step. 

\subsection{Reassembling principal component series} 
\label{sec:rs}
The spectral density of the $r$-dimensional principal component series $\X_t$'s obtained in (\ref{eq:brillinger}) is diagonal and aligned in descending order. However, the $k$-th assembled principal component series $X_{k,t}$ that is constructed using the eigenvector corresponding to the $k$-th largest eigenvalue at every frequency lack interpretability. 
As a matter of fact, given a pool of eigenvectors from different frequencies derived by previous steps, we can reassemble eigenvectors at different frequencies arbitrarily yet still maintain the zero coherence property of the resulting principal component series. The problem we are interested in solving is: given $(\vector v_{(1)}(\omega)$, $\vector v_{(2)}(\omega)$, $\cdots$, $\vector v_{(r)}(\omega))$, how do we assemble principal component series from a pool of eigenvectors at different frequences, so that to have an interpretable decomposition?

We build a solution to the above problem based on the fact that a shift in the time domain corresponds to a phase change in the frequency domain. This change is a linear function of frequency. For example, if a signal is propagating, the observations would take form as decayed and delayed versions of the original signal. Furthermore, if the rate of energy decay and the amount of time delay only depend on the location of observation and the original location of the signal, that is, does not depend on time, the corresponding phase of the Fourier transform of such signals would be perfectly correlated in the frequency domain. 

Simple dynamics such as signal propagating and sensing towards mobile energy sources lead to correlated phase in the frequency domain (see proofs in the Appendix). If we cluster eigenvectors with correlated phases to assemble principle component series, the filters constructed from each cluster would produce meaningful and interpretable spatiotemporal dynamics. 

\subsection{Decomposition by phase-aligned spectral filtering}
\label{sec:pasf}

In this section, we will outline the steps for the phase-aligned spectral filtering method. 

{\em Parameter estimation.} In practice, we only have observations on a finite time horizon. Let $\{z_{s,t}\in \R: s \in \D, t\in \T\}$ denote the observed data where $\T=\{1, \cdots, n\}\subset \mathcal T$. In order to assemble the desired phase-aligned spectral filters as described above, we need to estimate the paired filters $\{\C_{\tau} = (\C_{1,\tau}^\top, \cdots, \C_{r,\tau}^\top)^\top\in\R^{r\times m} \}$ and  $\{\B_{\tau}=(\B_{1,\tau}, \cdots, \B_{r,\tau})\in\R^{m\times r}\}$ from this finite sample so that the principal component series
$$
\x_{k,t} = \sum^{n}_{\tau=1} \C_{k, t-\tau} \z_{\tau}
$$
has diagonal or approximately diagonal spectral density and its resulting dynamic component $\fil{\z}^{(k)}_{t}$ obtained from 
$$ 
\fil{\z}^{(k)}_{t} = \sum^{n}_{\tau=1} \B_{k, t-\tau} \x_{k, \tau}
$$
has correlated phases across frequencies where $1\leq k\leq r$, $\C_{k, \tau}$ is the $k$-th row vector of $\C_\tau$ and $\B_{k, \tau}$ is the $k$-th column vector of $\B_\tau$.
 
The paired filters are constructed from the eigenvectors of spectral density matrices $f_{zz}(\omega)$. We estimate the eigenvectors $\vector{v}_{j}(\omega)$ by the eigenvectors of estimated spectral density $\hat{f}_{zz}(\omega)$. The spectral density matrices at Fourier frequencies $\omega_j=j/n$ for $j=0,\cdots, n-1$ are estimated by the smoothed periodogram
$$
\hat{f}_{zz}(\omega_{j}) = \sum^{q}_{k=-q}h_{k}P(\omega_{j}+\frac{k}{n}),
$$
where $P(\omega_{j})$ is the raw periodogram and $h_{k}$ is a smoothing kernel of bandwidth equal to $(2q+1)$ satisfying: i) $h_{k}>0$; ii) $\sum^{q}_{k=-q}h_{k}=1$; and iii) $q\rightarrow\infty$ and $q/n\rightarrow 0$, $\sum^{q}_{k=-q}h^{2}_{k}\rightarrow 0$ as $n\rightarrow\infty$.

{\em Phase unwrapping.}
We obtain the raw phase of the estimated $k$th eigenvector, $\Arg(\hat{\vector{v}}_{k}(\omega_j))$ at frequency $\omega_j$, by taking logarithm of the estimated eigenvector $\hat{\vector{v}}_{k}(\omega_j)$ and extract the imaginary part. However, the resulting phase estimate is only given as the actual phase modulo $2\pi$, between $-\pi$ and $\pi$. Even when the phase vectors of two eigenvectors are completely correlated, such a loss of information will render them much less correlated. In order to carry out our phase-aligned reassembling of eigenvectors, we need to recover the true phase up to a linear transformation. 

The computed raw phase (modulo $2\pi$) has discontinuities near $\pi$ and $-\pi$. We assume the true phase is continuous in space. Based on this assumption, we can then unwrap the raw phase values, in other words, resolve the jumps of phase values in a two dimensional space to derive continuous phase values. Over the spatial locations $s_k\in\R^2$, we apply the two-dimensional phase unwrapping algorithm proposed in~\cite{herraez2002fast} to each $\Arg(\hat{\vector{v}}_{k}(\omega))$ for $k=1,\cdots,r_\omega$.  The algorithm changes the raw values of $\Arg(\vector{v}_{k}(\omega_j))$ by adding $2c\pi$ with $c\in\mathbb Z$ at jumps so that the unwrapped phase values, denoted by $\widetilde{\Arg}(\hat{\vector{v}}_{k}(\omega_j))$ attain a maximum level of continuity over spatial locations. 

{\em Phase clustering.}
On the unwrapped phases $\widetilde{\Arg}(\hat{\vector{v}}_{k}(\omega_j))$ of the selected eigenvectors whose eigenvalues $\hat{\lambda}_{k}(\omega_j)$ are greater than or equal to $\Delta$, we deploy hierarchical clustering as the clustering algorithm with one minus correlation as the distance measure and Ward's clustering criterion~\cite{ward1963hierarchical} as the linkage agglomeration method. 

We then construct filters from each cluster to create reassembled principal component series that correspond to dynamics with correlated phases in their spectral densities. To construct the desired filters from each cluster, we label each of the selected eigenvector by its group number from the unwrapped phase clustering results. And then the paired filters for the $k$-th principal component series and its corresponding dynamic component are constructed by
\begin{align}
\hat{\C}_{k, \tau} &=\frac{1}{n} \sum^{n-1}_{j=0}\overline{\hat{\vector{v}}_k (\omega_{j})^\top}\I_{\hat{\lambda}_k (\omega_{j})\geq\Delta} \exp(2\pi i\tau\omega_{j}), 
\\
\hat{\B}_{k, \tau} &=\frac{1}{n} \sum^{n-1}_{j=0} \hat{\vector{v}}_k(\omega_{j})\I_{\hat{\lambda}_k (\omega_{j})\geq\Delta} \exp(2\pi i\tau\omega_{j}), 
\end{align}
where $\hat{\vector{v}}_k(\omega_j)$ is the eigenvector with a group label equal to $k$ when its eigenvalue $\hat{\lambda}_k (\omega_{j})$ is greater than or equal to $\Delta$.

{\em The phase-aligned spectral filtering algorithm. }
The complete phase-aligned spectral filtering decomposition procedure for spatiotemporal dynamics is summarized in Algorithm~\ref{algo:cluster}. The step numbers are the same as in Figure~\ref{fig:algorithm}.

\begin{algorithm}[H]
\SetKwInput{KwIn}{Input}
\SetKwInput{KwOut}{Output}
 \KwIn{Data $\{z_{s,t}: s\in\D, t\in\T\}$, number of top eigenvalues considered $r$ and threshold $\Delta$}
 \KwOut{dynamic components $\{\fil{z}^{(k)}_{s,t}: s\in\D, t\in\T, k=1,\cdots, r\}$}
 \caption{Phase-aligned spectral filtering decomposition}
 \label{algo:cluster}
  Estimate the spectral density $f_{zz}(\omega_j)$ for $\omega_j=j/n$ with $j=0,\cdots, n-1$; \\
  Calculate the top $r$ eigenvalues and eigenvectors for each $f_{zz}(\omega_j)$; \\
  Shrink the eigenvalues by the threshold $\Delta$ and obtain the correponding eigenvectors;\\
  Unwrap the phases of the selected eigenvectors; \\
  Cluster the selected eigenvectors using one minus the correlations of their unwrapped phases as dissimilarity; \\
  Construct paired filters from each cluster; \\
  Apply each of the $\C$ filters to the data $z_{s_{1}:s_{m},1:n}$ and obtain the reassembled principal component series; \\
  Apply each of the $\B$ filters to its corresponding principal component series and get the phase-aligned dynamic component.
    
\end{algorithm}

\section{Simulation results} 

We first illustrate the proposed phase-aligned spectral filtering method using multiple constructed low-dimensional dynamic systems in an area where observations of the entire system are taken on a $20\times 20$ grid of spatial locations: on $\mathcal D=[0,20]^2\subset \R^2$ with grid blocks $\{s_{j,k}=[j-1,j]\times [k-1,k]\subset \R^2: 1\leq j\leq 20, 1\leq k\leq 20, j\in\mathbb N, k\in\mathbb N\}$ and grid locations being the centers of the corresponding grid blocks. The phase-aligned spectral filtering method is then compared with a number of comparison methods found in the literature. 

\subsection{Scenario I: rotating energy sources}
In this example, we create a scenario where the observed value at a given grid location and a given time point is the total energy absorbed by the unit block area centered at this given grid location, from all rotating energy sources.

On the $20\times 20$ grid, there are two energy sources affecting the area, each of which moves following a circular trajectory. The two trajectories are centered at $c^{(1)}_c = (15,15)$ and $c_c^{(2)} = (5,5)$ respectively, with radius of $r^{(1)}_c=r^{(2)}_c=5$. The two energy sources move different angular velocities of $v^{(1)}_{\theta} = 2\pi/20$ and $v^{(2)}_{\theta} = 2\pi/5$ per time unit counterclockwise respectively. The initial positions of the two energy sources on the trajectory circles, denoted by $\theta^{(1)}_0$ and $\theta^{(2)}_0$ measuring the angular distance from the horizontal axis. They are randomly assigned in each simulation. The $i$-th energy source's position at $t=0$ can therefore be written as $c^{(i)}_c + (r^{(i)}_c\cos\theta^{(i)}_0, r^{(i)}_c\sin\theta^{(i)}_0)$ with $\theta^{(i)}_0$ uniformly sampled between $0$ to $2\pi$ for $i=1,2$. At any subsequent time point $t$, the $i$-th energy source's position is
$$
c^{(i)}_t=\binom{c^{(i)}_{t,1}}{c^{(i)}_{t,2}} = c^{(i)}_c + \binom{r^{(i)}_c\cos(\theta^{(i)}_0+v^{(i)}_\theta t)}{r^{(i)}_c\sin(\theta^{(i)}_0+v^{(i)}_\theta t)}, \,\, i=1,2.
$$

\begin{figure}[t]
	\centering
	\captionsetup[subfigure]{position=top}
	\subfloat[][Low-noise: $\sigma^2_\varepsilon=0.16$]{
		\includegraphics[width=0.33\textwidth]{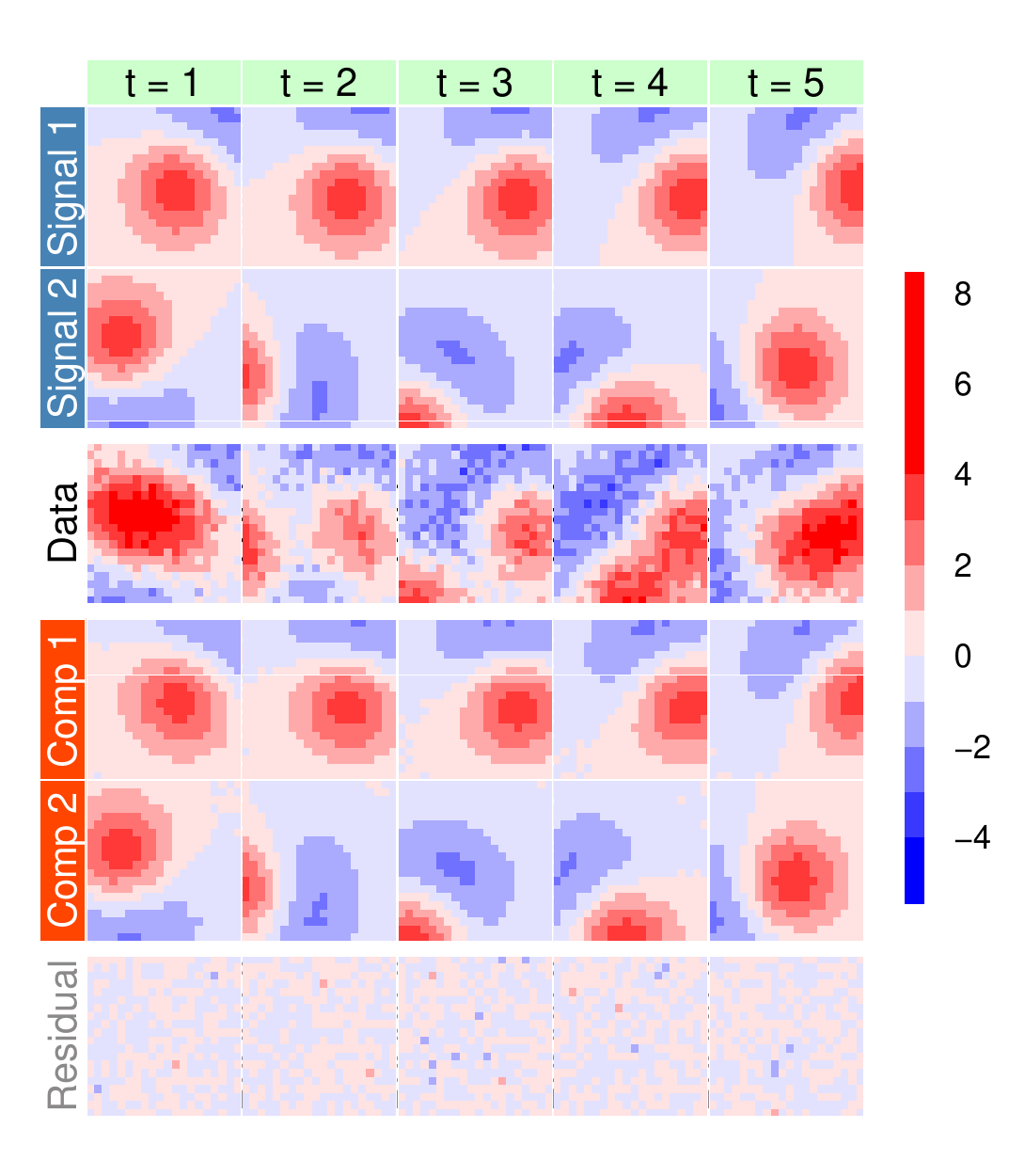}}
	\subfloat[][Mid-noise: $\sigma^2_\varepsilon=4$]{
		\includegraphics[width=0.33\textwidth]{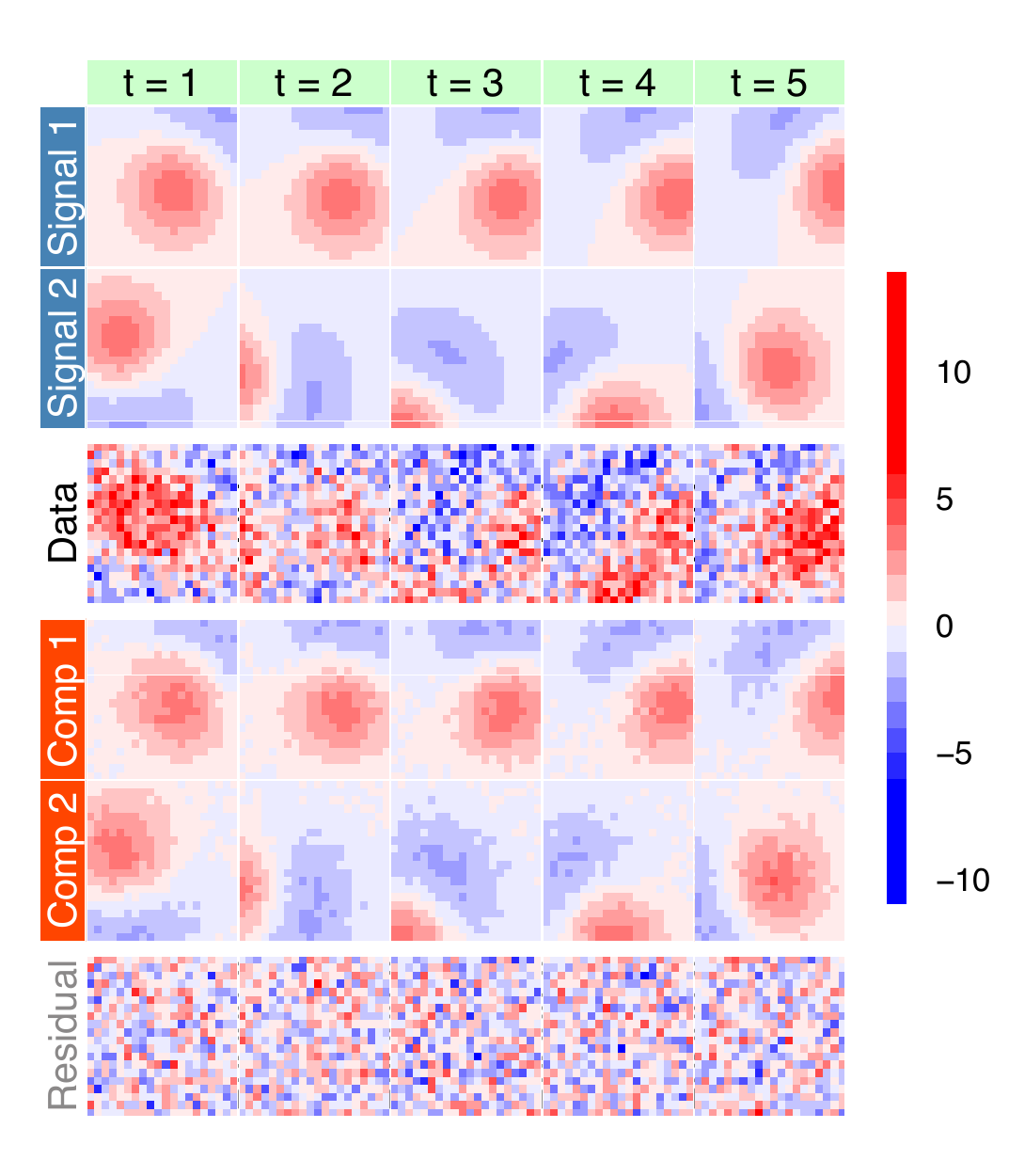}}
	\subfloat[][High-noise: $\sigma^2_\varepsilon=16$]{
		\includegraphics[width=0.33\textwidth]{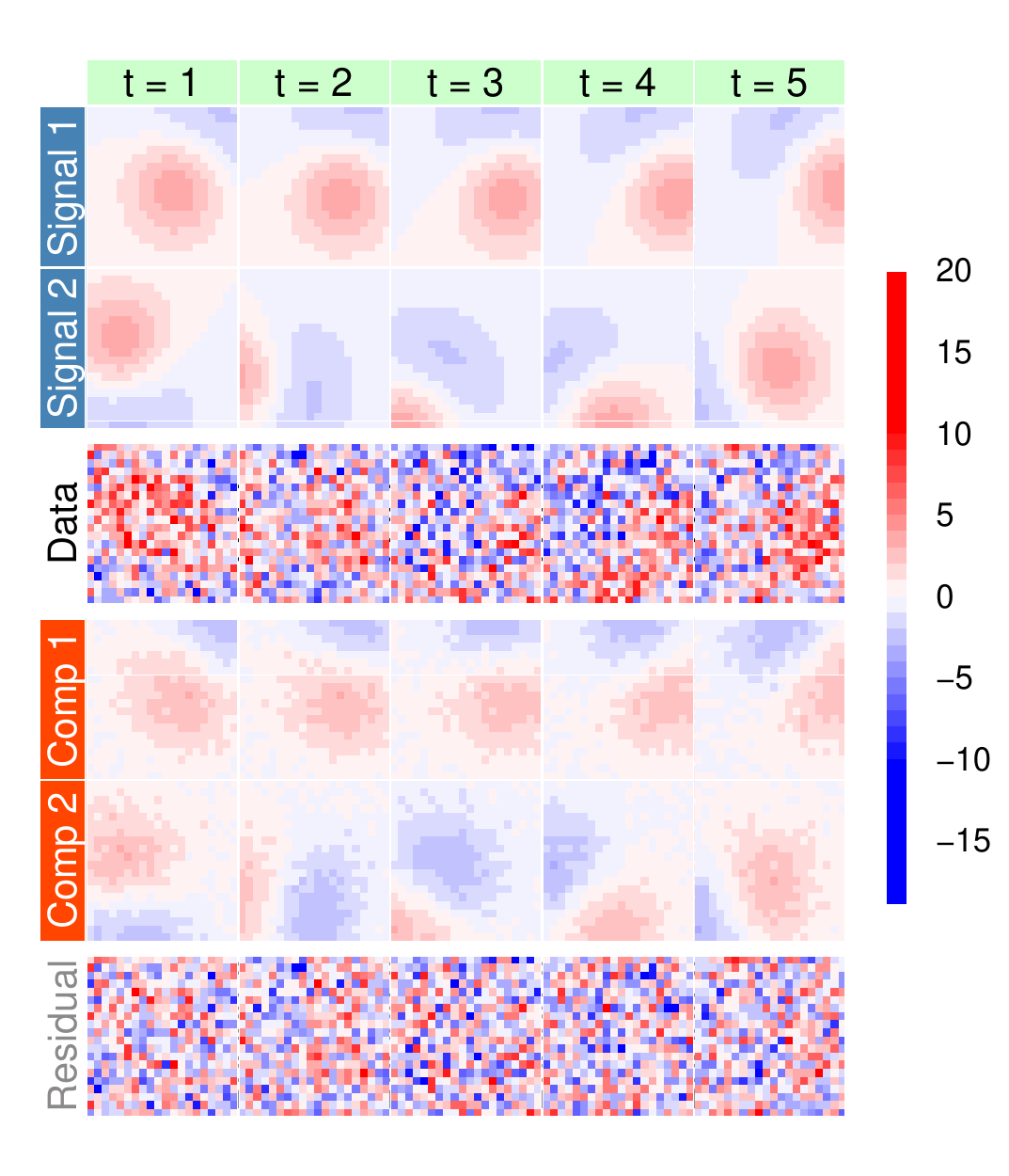}\label{fig:sim.twoball.cluster.high}}
	\caption{Phase-aligned dynamic decomposition in the two-rotating-energy-source simulation from $t=1$ to $t=5$ in the low-, mid- and high-noise level settings.}
	\label{fig:sim.twoball.cluster}
\end{figure}

Assume that the energy absorbed from the energy source decays exponentially in squared distance. For $s\in \mathcal D, t\in\T=\{1,2,\cdots,n\}$, we can explicitly write down the energy at location $s$ and time $t$ absorbed from the $i$-th energy sources positioned at $c^{(i)}_t$ as
$$
E^{(i)}_{s,t} = E^{(i)}_0\exp\Big(-\frac{\|s-c^{(i)}_t\|^2}{\gamma^{(i)}}\Big)
$$
where $E^{(i)}_0$ is the total emitted energy of the $i$-th energy source during any unit time. $E^{(i)}_0$ is assumed to be a constant for simplicity. The total amount of energy measured at location $s$ and time $t$ is the sum of energy absorbed from all sources, that is, 
$$
E_{s,t} = \sum_{i=1}^{2}E^{(i)}_{s,t} = \sum_{i=1}^{2}E^{(i)}_0\exp\Big(-\frac{\|s-c^{(i)}_t\|^2}{\gamma^{(i)}}\Big).
$$ 
Thus, the amount of energy the grid block $s_{j,k}$ absorbs at time $t$ is
$$
 z_{(j,k),t} = \int_{s\in s_{j,k}} E_{s,t} ds.
 $$
%
%&= \int_{j-1}^{j}\int_{k-1}^{k}\sum_{i=1}^{2}E^{(i)}_0\exp\Big(-\frac{(x_1-c^{(i)}_{t,1})^2+(x_2-c^{(i)}_{t,2})^2}{\gamma^{(i)}}\Big) dx_1dx_2. \\
%
%We can write $z_{(j,k),t} = \sum_{i=1}^{2}z^{(i)}_{(j,k),t}$, where
%\begin{align*}
%&z^{(i)}_{(j,k),t} \\
%& = \int_{j-1}^{j}\int_{k-1}^{k}E^{(i)}_0\exp\Big(-\frac{(x_1-c^{(i)}_{t,1})^2+(x_2-c^{(i)}_{t,2})^2}{\gamma^{(i)}}\Big) dx_1dx_2.
%\end{align*}
The total emitted energy $E^{(i)}_0$ is set to be $1000$ and the bandwidth parameter $\gamma^{(i)}$ is set to be $5$ for $i=1,2$. We use the demeaned $\z^{(1)}_t$ and $\z^{(2)}_t$ as the underlying dynamic systems that affect the grid area. The final observed measurements are the energy distributed by these two rotating sources overlaid and superimposed on each other with high dimensional white noises added, i.e., 
$$
y_{(j,k),t}= z_{(j,k),t} + \varepsilon_{(j,k),t}, \,\, 1\leq j\leq 20, 1\leq k\leq 20.
$$
Using vectorized notation for $\varepsilon_{(j,k),t}$,  we define $\vector{\varepsilon}_t=(\varepsilon_{1,t}, \varepsilon_{2,t}, \cdots, \varepsilon_{400,t})^\top$. We assume that $\vector{\varepsilon}_t\sim\mathcal N(0,\sigma^2_\varepsilon\mathbf I_m)$ where $m=400$. We set three noise levels in this simulated  scenario: low-noise level with $\sigma^2_\varepsilon=0.16$, mid-noise level with $\sigma^2_\varepsilon=4$ and high-noise level with $\sigma^2_\varepsilon=16$. The top three rows of Fig.~\ref{fig:sim.twoball.cluster} are the level plots of the two rotating energy sources $\z^{(1)}_t$ and $\z^{(2)}_t$ along with the observed data $\vector{y}_t$ from $t=1$ to $t=5$ under the three noise-level settings. As one can see, when $\sigma^2_\varepsilon$ increases to $16$, the two dynamic systems are barely discernible in the superimposed observed data.
%
%\begin{figure}[htbp]
%	\centering
%	\captionsetup[subfigure]{position=top}
%	\includegraphics[width=0.5\textwidth]{twoball_data.pdf}
%	\caption{Level plots of the two rotating energy sources and observed data at selected time in the low-noise level setting with $\sigma^2_\varepsilon=0.16$}
%	\label{fig:sim.twoball.data}
%\end{figure}

The performance of our phase-aligned spectral filtering (PASF) algorithm was evaluated on the simulated data under these three noise level settings. See Appendix for implementation details. The lower three rows of Fig.~\ref{fig:sim.twoball.cluster} display the decomposition results by phase-aligned spectral filtering (PASF) in the low-, mid- and high-noise level settings respectively. Rows 4 and 5 are the identified dynamic components using our proposed method. Under the low-noise setting, the first and second components explain 48\% and 47\% of the variability in  observed data respectively. When $\sigma^2_\varepsilon=4$ (mid-noise level), each of the two dynamic components accounts for 23\% of the data's variability. When $\sigma^2_\varepsilon$ increases to 16 (an overwhelmingly high noise level), variability carried by the two filtered components drops to 8\%. It can be seen clearly from Fig.~\ref{fig:sim.twoball.cluster} that the phase-aligned spectral filtering (PASF) approach is still able to capture and separate the underlying dynamic systems even when the signal to noise ratio drops below $0.1$ (the variance of the signals is approximately $1.6$). Row 6 of each panel displays the residuals after we subtract the filtered components constructed by phase-aligned spectral filtering (PASF), which resemble white noises. See {\em Supplement Information} for animated plots of these simulation results. 

\begin{figure}
	\centering
	\includegraphics[width=\textwidth]{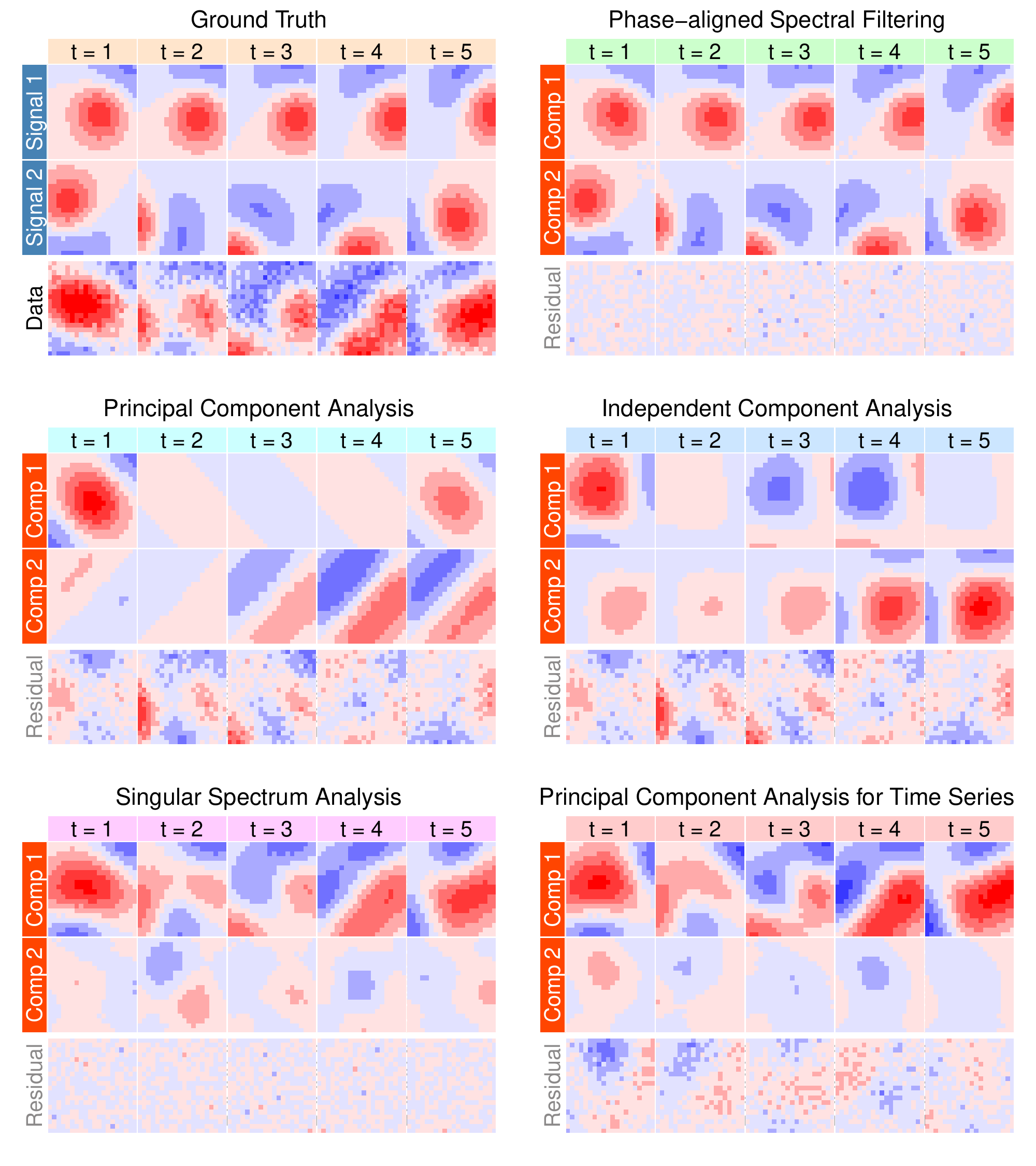}
	\caption{Decomposition results from different methods in the {\em two rotating energy sources} simulation example under the low-, mid- and high-noise level settings. \label{fig:sim.twoball.all}}
\end{figure}  
For comparison, we also applied the principal component analysis (PCA), independent component analysis (ICA), singular spectrum analysis (SSA)~\cite{ghil2002advanced} and principal component analysis for time series (PCA4TS)~\cite{chang2014segmenting} to the simulated data from the {\em rotating energy source} example. Fig.~\ref{fig:sim.twoball.all} provides a side-by-side comparison of the decomposition results obtained by phase-aligned spectral filtering (PASF) and these methods found in the literature, under the low-noise level setting with $\sigma^2_\varepsilon=0.16$. The top left panel displays the ground truth, i.e., the true signals $\z^{(1)}_t$ and $\z^{(2)}_t$ that generate the observed data $\vector{y}_t$. The top right panel is the two components as well the residuals from our phase-aligned spectral filtering (PASF) method. The remaining panels are the resulting dynamic components and corresponding residuals computed by the literature methods (See Appendix for implementation details of these literature methods.). Comparison results under the mid- and high-noise level setting can be found in Appendix as Fig.~\ref{fig:sim.twoball.all.mid} and Fig.~\ref{fig:sim.twoball.all.high}.
  
None of the literature methods is able to recover the two underlying dynamic systems as our method does. The dynamics yielded from PCA and ICA are distant from the true dynamics since they can only capture linear features with no time dependence. SSA is able to separate the smooth dynamics from the noisy data but failed to separate them. The performance of PCA4TS is better than PCA and ICA but worse than SSA in the low-noise setting, and behaves more like PCA in the mid- and high-noise level settings. 

Animated level plots of these decompositions can be found in Appendix. 

\subsection{Scenario II: signal propagation}

The second spatiotemporal system of dynamics considered in our simulation study involves a scenario where multiple signal processes propagate on a grid. Different from the previous scenario, here the signal sources do not move. Rather, the signals propagate along preset directions. Observed value at any spatial location and at a given time $t$ is then the sum of all propagated signals at this location and time. We further assume that the magnitude of signals decays as it propagates. Assume that four independent univariate autoregressive processes, denoted as $X_k$ for $k=1,2,3,4$, sit at the four corners of the grid. The observed value at grid block $s_j$ for $j=1,\cdots, 400$ at time $t$ can be written as the sum of four independent signal processes with different time lags:
$$
Y_{s_j,t} = \sum^{4}_{k=1} a_{jk}X_{k, t-t_{jk}},
$$
where $\{a_{jk}, k=1,\ldots, 4\}$ are linear weights at location $s_j$ for the signals $X_k$ and $\{t_{jk}, k=1,\ldots 4\}$ are lag delays between $s_j$ and $X_k$ respectively. Each signal, $\{X_{k,t}\}$, $k=1,\ldots,4$, is an autoregressive process of order 2, that is,
$$
X_{k,t} = \beta_{k,1} X_{k,t-1} + \beta_{k,2} X_{k,t-1} + \varepsilon_{k,t},\qquad \varepsilon_{k,t}\sim\mathcal N(0,1)
$$
with $\beta_{1,1}=\beta_{2,1}=0.9$, $\beta_{3,1}=\beta_{4,1}=-0.9$, $\beta_{1,2}=\beta_{3,2}=-0.5$, and $\beta_{2,2}=\beta_{4,2}=-0.8$.

%
%\begin{figure}
%	\centering
%	\includegraphics[width=0.8\textwidth]{foursignal_data.pdf}
%	\caption{Level plots of the four propagating signals and observed data at selected time points.}
%	\label{fig:sim.foursignal.data}
%\end{figure} 
%

Denote the location coordinates for $X_{k,t}$ by $c_{k}$. In our simulation, we use $c_1=(0,0)$, $c_2=(20,0)$, $c_3=(0,20)$ and $c_4=(20,20)$.
For $Y_{s_j,t}$, $a_{jk}$ and $t_{jk}$ are decided by the spatial distance between the grid block $s_j$ and the location of signals. Specifically,
$$
	a_{jk} = \exp(-\|s_j-c_k\|_2/\gamma), \,\, t_{jk} = \|s_j-c_k\|_1, 
$$
where $\|\cdot\|_1$ and $\|\cdot\|_2$ are the $L_1$ and $L_2$ norm respectively.
% defined as 
%$$
%	\|g\|_1 = \sum_{i=1}^{2}|g_i|,\,\|g\|_2 = \Big(\sum_{i=1}^{2}g_i^2\Big)^{1/2},
%$$
%for $g=(g_1,g_2)$ 
Parameter $\gamma$ is the signals' rate of decay when propagating and is set to be 50 in this simulation. The observed value $Y_{s_j, t}$ is then the sum of four dynamic components $Y_{s_j,t}^{(1)}$, $Y_{s_j,t}^{(2)}$, $Y_{s_j,t}^{(3)}$ and $Y_{s_j,t}^{(4)}$ where
$$
Y_{s_j,t}^{(k)} = e^{-\|s_j-c_k\|_2/\gamma} X_{k, t-\|s_j-c_k\|_1}, \,\, k=1, 2, 3, 4.
$$

\begin{figure}
	\centering
	\includegraphics[width=0.8\textwidth]{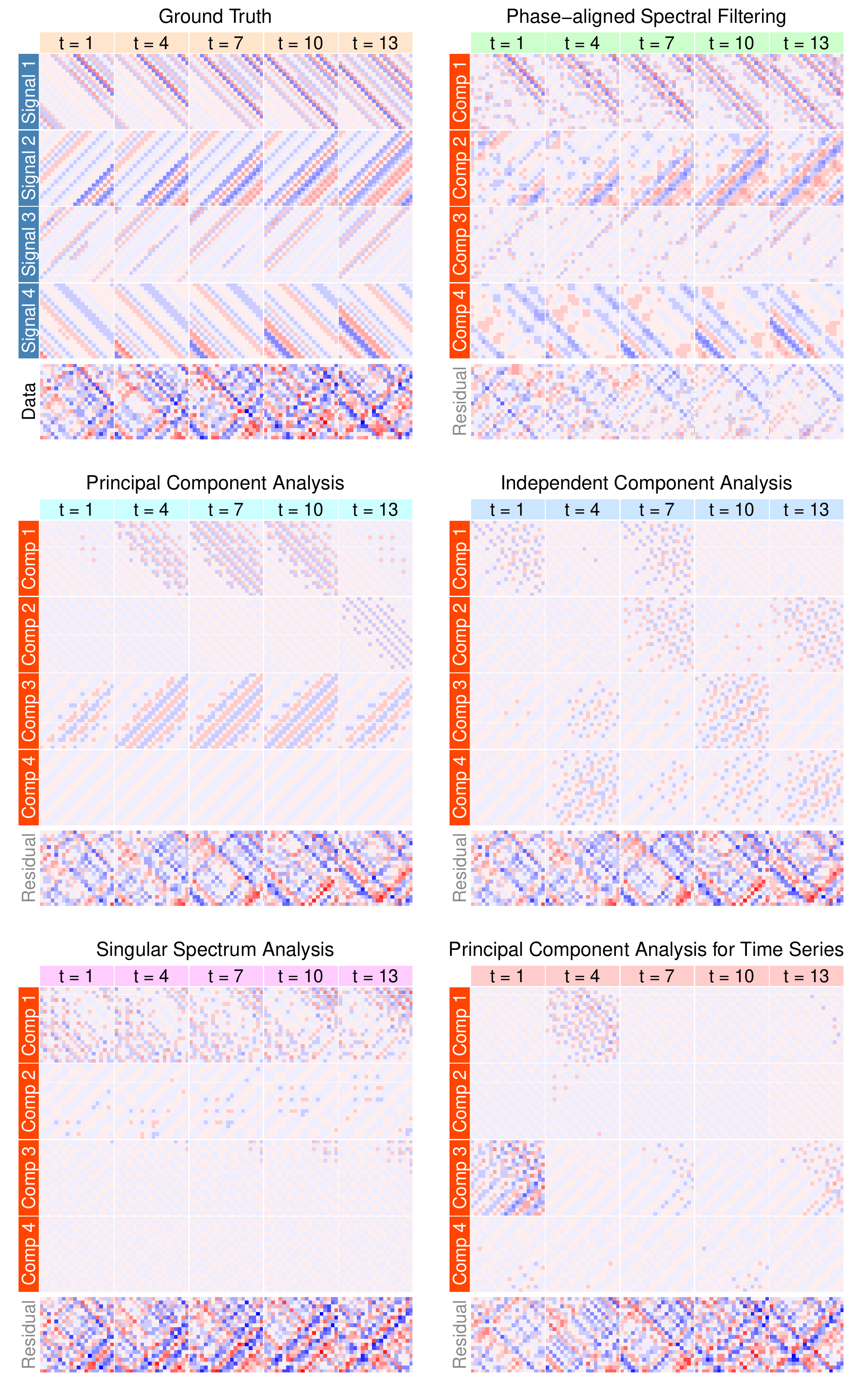}
	\caption{Decomposition results from different methods on data simulated in the four propagating signals  example.}
	\label{fig:sim.foursignal.all}
\end{figure}

Let the vectorized notation of the propagating signal over the grid at time $t$ be
$\vector{Y}_{t}^{(k)} = (Y_{s_{1},t}^{(k)}, Y_{s_{2},t}^{(k)}, \cdots, Y_{s_{400},t}^{(k)})^\top$. 
The top-left panel of Figure~\ref{fig:sim.foursignal.all} displays the true propagating components  $\vector{Y}_{t}^{(1)}$, $\vector{Y}_{t}^{(2)}$, $\vector{Y}_{t}^{(3)}$ and $\vector{Y}_{t}^{(4)}$ along with the aggregated signals as observed data. The estimated variances of the four signals are 2.55, 2.02, 1.14, and 1.09. The top-right panel of Figure~\ref{fig:sim.foursignal.all} displays the dynamic components obtained from the proposed phase-aligned spectral filtering (PASF) method and the corresponding residuals. The corresponding components resulted from our method account for 36\%, 31\%, 16\% and 14\% of all variability in the observed data.  Although the data do not display any evident patterns or dynamics, our proposed approach still manage to detect and separate the four propagating signals from the very noise-like data. 

We also compare our approach with the same four literature methods that were used in the previous example. Figure~\ref{fig:sim.foursignal.all} shows the decomposition results from these methods for comparison. The animated version of these results can be found in the Appendix. As we can see, PCA identifies the direction orthogonal to the direction of signal propagating but fails to capture the dynamics. Other literature methods capture even less than PCA. In this example, only our phase-aligned spectral filtering (PASF) method can almost fully recover the dynamic components corresponding to the four propagating signals. 

\section{Real Data Analysis}
The climate data analyzed in this paper is daily sea level pressure from NCEP/NCAR Reanalysis~\cite{kalnay1996ncep}. The data has a spatial resolution of $2.5^{\circ}$ latitude $\times$ $2.5^{\circ}$ longitude. The grid covers a part of the pacific ocean from $30^{\circ} $N to $60^{\circ} $N and from $150^{\circ} $E to $230^{\circ} $E, which corresponds to a total of 429 spatial locations. We used observations from April 6th, 2012 to December 31, 2014, a total of 1000 time points. 

\begin{figure}
	\centering
	\includegraphics[width=\textwidth]{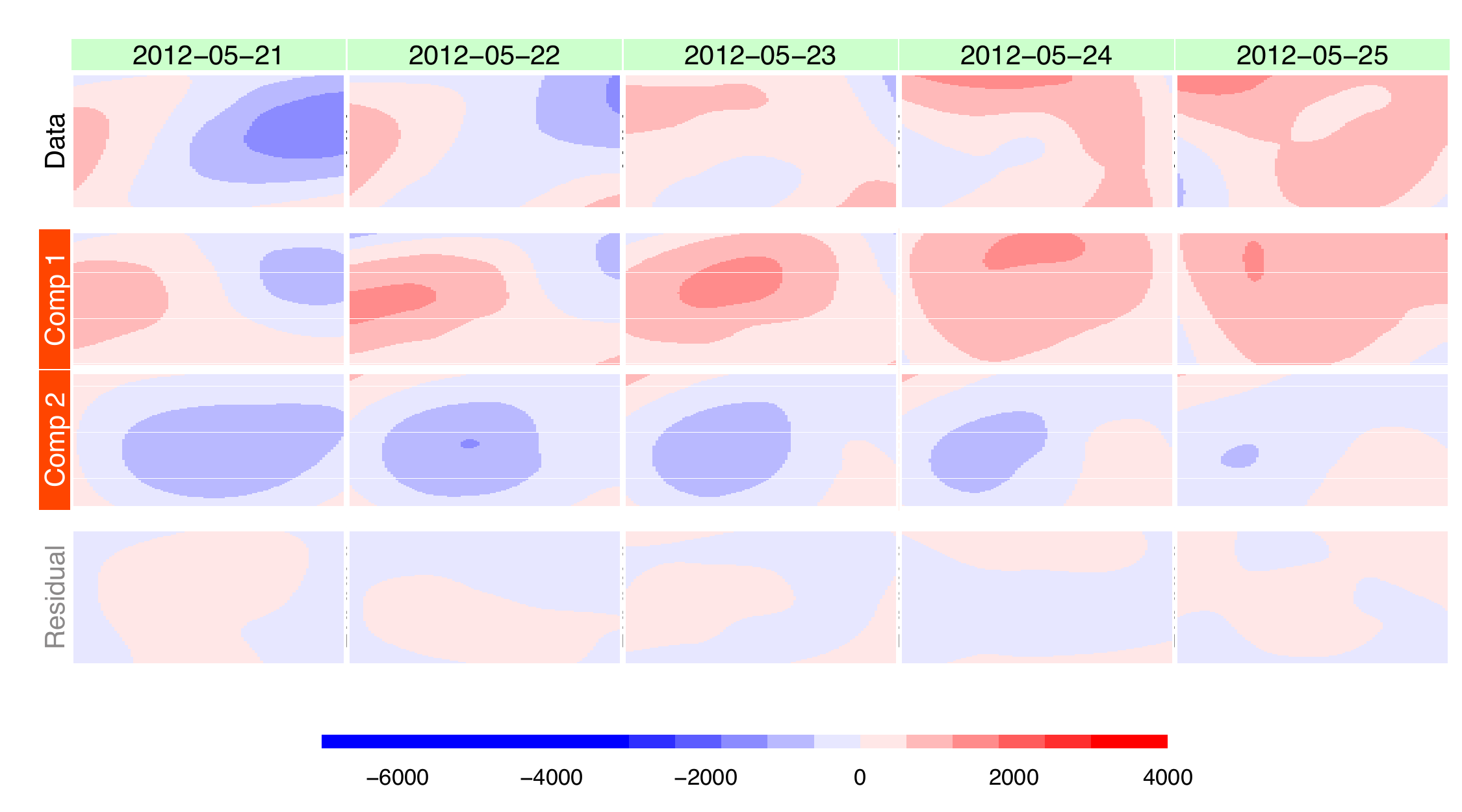}
	\caption{Phase-aligned dynamic decomposition of the sea level pressure data.}
	\label{fig:slp}
\end{figure}

Two dynamic components were identified by the proposed method.  The dynamic component obtained from each cluster accounts for 63\% and 32\% of the total variability respectively. Figure~\ref{fig:slp} shows the two dynamic components obtained from phase-aligned spectral filtering (PASF) as well as the observed data and the residuals after spectral filtering for 5 days, May 21, 2012 to May 25, 2012. During the time range displayed in Figure~\ref{fig:slp}, the first component captures a high level pressure dynamic moving from west to east and the second component captures a low level pressure dynamic moving from east to west. The animated level plots in the Appendix show similar trends throughout the entire time range. It can be seen from our decomposition results that the first component describes processes generate from the west side and propagate to the east while the second component captures processes generate from the east side and propagate to the west. These two dynamic components obtained from our approach explain a total of 95\% of the information carried by the observed data.

\begin{figure}
	\centering
	\includegraphics[width=\textwidth]{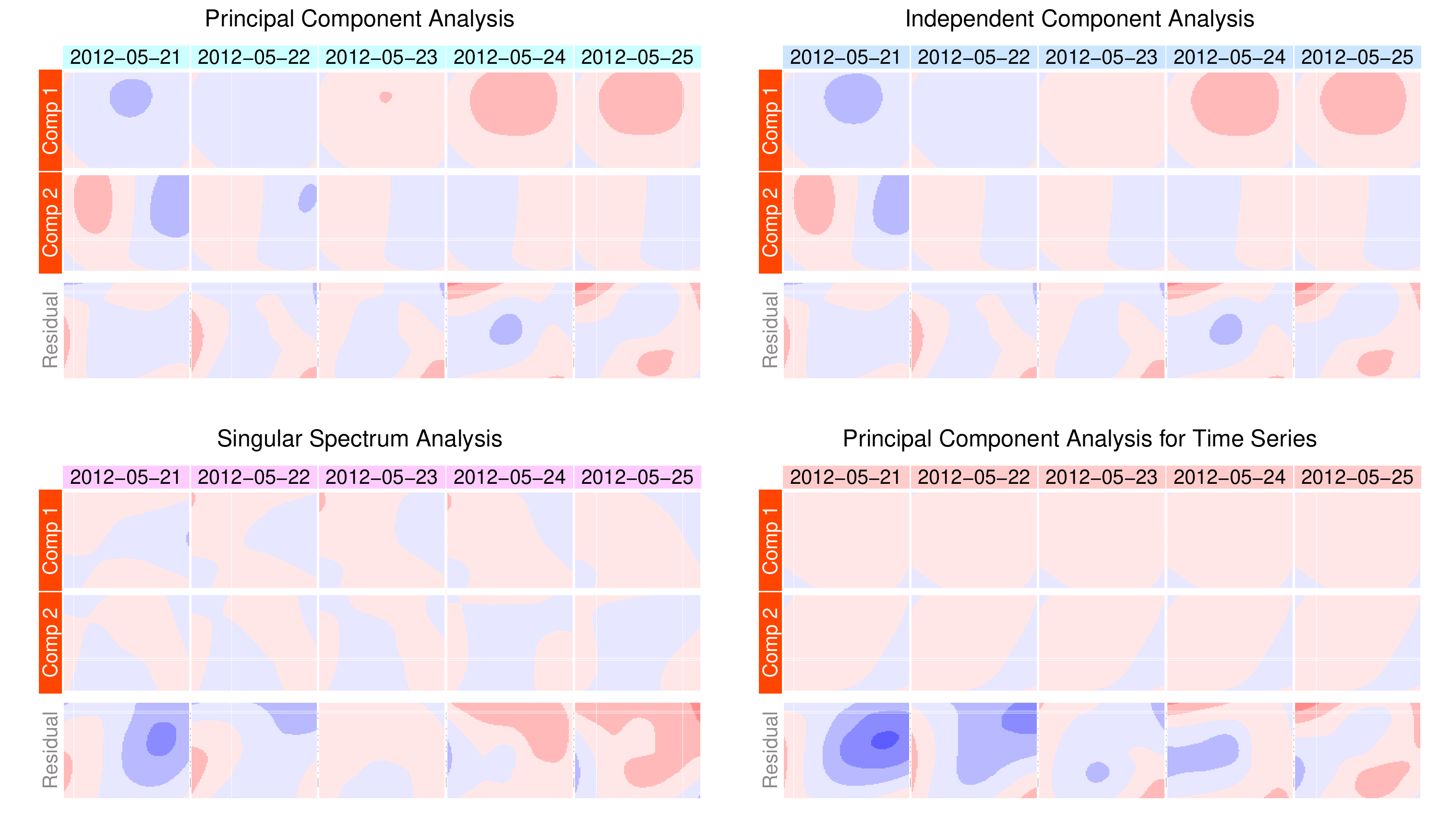}
	\caption{Decomposition results from different literature methods on the sea level pressure data.}
	\label{fig:slp.ref}
\end{figure}

We also applied PCA, ICA, SSA and PCA4TS to this real dataset. The resultant components from these four methods all explain about 48\% of variability in the data.  Figure~\ref{fig:slp.ref} shows the decomposition results as well the residuals from these literature methods from May 21, 2012 to May 25, 2012. The components are ordered by their variances. There are no evident dynamic patterns in the components extracted by these methods. Furthermore, the residuals still carry visible information and dynamics. 

\section{Discussion}

In this paper, we propose a phase-based clustering method to create interpretable components that correspond to low-rank spatiotemporal dynamic signals with correlated phase across frequencies. In the two simulated scenarios of multiple signal sources propagating or rotating spatially, our method demonstrates excellent capabilities of capturing and separating the underlying low-dimensional dynamic systems. We also obtain interesting patterns from the components extracted using our algorithm from the analysis of sea level pressure data. The class of dynamics with phase-aligned spectral density could be very rich. Our results show that this class of dynamics include signal propagation and energy resource rotating on periodic curves (see Appendix). Although we have not fully understood its full geometric structure, our algorithm obtained clean and interpretable lower rank spatiotemporal dynamics that explains a substantial proportion of the observed data in both our simulation study and analysis of climate data. Especially our approach outperforms other methods found in the literature, in terms of both information retrieval and signal separation. 

\bibliographystyle{unsrt}
\bibliography{refs}

\newpage
\appendix

\setcounter{figure}{0}
\renewcommand\thefigure{A\arabic{figure}}    
\renewcommand\thesection{A\arabic{section}}    

\section{Computation details for examples}

\paragraph{Rotating energy source examples.} For applying phase-aligned spectral filtering (PASF) to the {\em rotating energy source} examples, the bandwidth of the Daniell kernel used for smoothing the raw periodogram is chosen to be $21$. The threshold $\Delta$ is decided by the spectral gap in the pooled eigenvalues. The total number of selected eigenvectors is $80$ for the low- and mid-noise cases and $38$ for the high-noise case. 

For PCA and ICA, we extract the first two components with highest energy, that is, the two components corresponding to the first two largest eigenvalues. 

For SSA, the first 10 largest singular values are dominant in the low- and mid-noise level settings. In the high-noise setting, the gap appears between the eighth and ninth largest singular values. We divide the corresponding components into two groups by clustering them based on their weighted correlation and reconstruct the final two components from each group. The weighted correlation $\rho_w(\cdot, \cdot)$ is a measure of the degree of separability between two series and defined as
$$
\rho_w(\zeta_t, \eta_t) = \frac{(\zeta_t, \eta_t)_w}{\|\zeta_t|_w\|\eta_t\|_w},
$$
where $\|\cdot\|_w = \sqrt{(\cdot, \cdot)_w}$, $(\zeta_t, \eta_t)_w = \sum_{k=1}^{n}w_k\zeta_k\eta_k$, $w_k=\min(k, n/2, n-k)$, and both of the two series $\zeta_t$ and $\eta_t$ have a time length equal to $n$. 

For PCA4TS, we choose the two groups whose corresponding components in the original space have largest variance.

\paragraph{Signal propagation examples.} The bandwidth of the Daniell kernel for smoothing the raw periodogram is chosen to be $21$. The total number of selected eigenvectors is 1000 and the dendrogram displays four clean clusters. The resulting components obtained from our algorithm show evident dynamics of four processes propagating on the grid area which perfectly correspond to the generated signals.

We extract the first four components for the PCA and ICA respectively. For SSA, we divide the first 50 components into four groups by clustering them based on their weighted correlation and reconstruct one new component from each group. For PCA4TS, we choose the four groups whose corresponding components in the original space have the largest variance.

\paragraph{Real data analysis: sea level pressure.} The observations were first demeaned for each location before analysis. The bandwidth of the Daniell kernel for smoothing the periodogram was chosen to be 21. The total number of selected eigenvectors is 2000 and the resulting number of clusters is 2. 

As in the simulation examples, we extract the first two components from PCA and ICA. We cluster the components from SSA into two groups based on their weighted correlation to obtain two final additive components. We select the first two components with the largest variance from PCA4TS.

\section{Connecting phase correlation with simple spatiotemporal dynamics}
In this section we show that two common spatiotemporal dynamics can create phase correlation between their spectral densities at different frequencies. 

\subsection{Signal Propagation}
Assume that there are $r$ independent real-valued signal processes $\{X_{1,t}, \cdots, X_{r,t}\}$ propagating and the observed process $\Z_t = (Z_{s_1,t}, \cdots, Z_{s_m,t})^\top\in\R^m$ satisfies
$$
Z_{s_j,t} = \sum_{k=1}^{r}a_{jk} X_{k,t-t_{jk}},
$$
where $a_{jk}>0$ and the spectral density of $(X_{1,t}, X_{2,t}, \cdots, X_{r,t})^\top$ is
$$
f_{xx}(\omega) = 
\begin{pmatrix}
\lambda_1(\omega) & & \\
& \ddots & \\
& & \lambda_r(\omega) \\
\end{pmatrix}.
$$
Then the element of the spectral density of $\Z$ in the $j$-th row and $l$-th column is given by
$$
[f_{zz}(\omega)]_{j,l} = \sum_{k=1}^r a_{jk}a_{lk}  e^{-2\pi i\omega (t_{jk} - t_{lk})} \lambda_k(\omega).
$$	
If we define
\begin{align*}
\mathbf A(\omega) & = [a_{jk}e^{-2\pi i\omega t_{jk}}]_{1\leq j\leq m, 1\leq k\leq r} \\
&=
\begin{pmatrix}
a_{11}e^{-2\pi i\omega t_{11}}& a_{12}e^{-2\pi i\omega t_{12}}& \cdots & a_{1r}e^{-2\pi i\omega t_{1r}} \\
a_{21}e^{-2\pi i\omega t_{21}}& a_{22}e^{-2\pi i\omega t_{22}}& \cdots & a_{2r}e^{-2\pi i\omega t_{2r}} \\
\vdots & \vdots & \cdots & \vdots \\
a_{m1}e^{-2\pi i\omega t_{m1}}& a_{m2}e^{-2\pi i\omega t_{m2}}& \cdots & a_{mr}e^{-2\pi i\omega t_{mr}} \\	
\end{pmatrix}
\end{align*}
then $f_{zz}(\omega) = \mathbf A(\omega) f_{xx}(\omega) \overline{\mathbf A(\omega)^\top}$. The phase of $\mathbf A(\omega)$ is $[-2\pi\omega t_{jk}]_{1\leq j\leq m, 1\leq k\leq r}$ which is a linear function of $\omega$.

\subsection{Rotating Energy Source}
Consider a mobile energy source defined by $(c_t, E_t)$ where $c_t\in \R^2$ is the rotating trajectory and $E_t\in\R$ is the energy it carries at time $t$. Assume that $c_t$ orbits around a center $c_0$ with uniform angular speed $v_\theta$. That is,
$$
c_t = c_0 + \binom{r\cos(\theta_0+v_\theta t)}{r\sin(\theta_0+v_\theta t)}.
$$
where $\theta_0$ is the initial angle and $r$ is the distance between $c_0$ and $c_t$.

Now consider the observation $z_{s,t}$ being the absorbed energy from the signal at location $s$ and time $t$ where $s\in\R^2$ can be written as
$$
s = c_0 + \binom{r_s\cos\theta_s}{r_s\sin\theta_s}.
$$
Assume that $z_{s,t}$ is in the form of 
$$
z_{s,t} = E_t\cdot\kappa(\|s-c_t\|^2),
$$
where $\kappa(\cdot)$ is a non-negative real-valued monotone decreasing function satisfying $\kappa(0) \leq 1$, and $\|s-c_t\|$ is the Euclidean distance between $s$ and $c_t$, that is,
\begin{align*}
\|s-c_t\|^2 &= r^2 + r_s^2 - 2rr_s\cos(\theta_0+v_\theta t - \theta_s) \\
&= r^2 + r_s^2 - 2rr_s\cos\left(v_\theta \Big(t - \frac{\theta_s-\theta_0}{v_\theta}\Big)\right).
\end{align*}
We assume $E_t = E_0$ for stationarity, that is,  $E_t$ does not change over time.
Let
$$
f_{s,t} = E_0\cdot\kappa(r^2+r_s^2 - 2rr_s\cos(v_\theta t))
$$
and $\mathcal F_{s}$ be the Fourier Transform of $f_{s,t}$. 
Then the Fourier Transform of $z_{s,t}$ is 
$$
\mathcal F_{z_{s}} (\omega) = F_{s}(\omega) \exp\Big(-2\pi i \omega \frac{\theta_s-\theta_0}{v_\theta}\Big)
$$
Since $f_{s,t}$ is a symmetric function of $t$, $\mathcal{F}_s$ is real.
Therefore the modulus of $\mathcal F_{z_s}$ is $\mathcal |F_{s}(\omega)|$
and the phase
$$
\mbox{Arg}(\mathcal F_{s} (\omega)) = \left\{
\begin{array}{ll}
-2\pi\omega \frac{\theta_s-\theta_0}{v_\theta} & \mbox{if}\quad F_{s}(\omega) \geq 0 \\
-2\pi\omega \frac{\theta_s-\theta_0}{v_\theta} + \pi& \mbox{if}\quad F_{s}(\omega) < 0
\end{array} 
\right.
$$
is a linear function of $\omega$.

\section{Additional figures}
\begin{table}[htbp]
	\caption{Links to Animated Figures}
	\centering
	\begin{tabular}{l|l}
		\hline
		URL & Description\\
		\hline
		\url{http://goo.gl/LePZNs} & two rotating energy sources, simulation example, low-noise level\\
		\url{http://goo.gl/tGld5F} & two rotating energy sources, simulation example, mid-noise level\\
		\url{http://goo.gl/YB9HcW} & two rotating energy sources, simulation example, high-noise level\\
		\url{http://goo.gl/JCpAjB} & four propagating signals, simulation example\\
		\url{http://goo.gl/wiKEVt} & sea level pressure, real data example\\
		\hline
	\end{tabular}
	\label{table:animation}
\end{table}

\begin{figure}[hb!]
	\centering
	\captionsetup[subfigure]{position=top}
	\subfloat[][Low-noise level: $\sigma^2_\varepsilon=0.16$]{
	\includegraphics[width=0.8\textwidth]{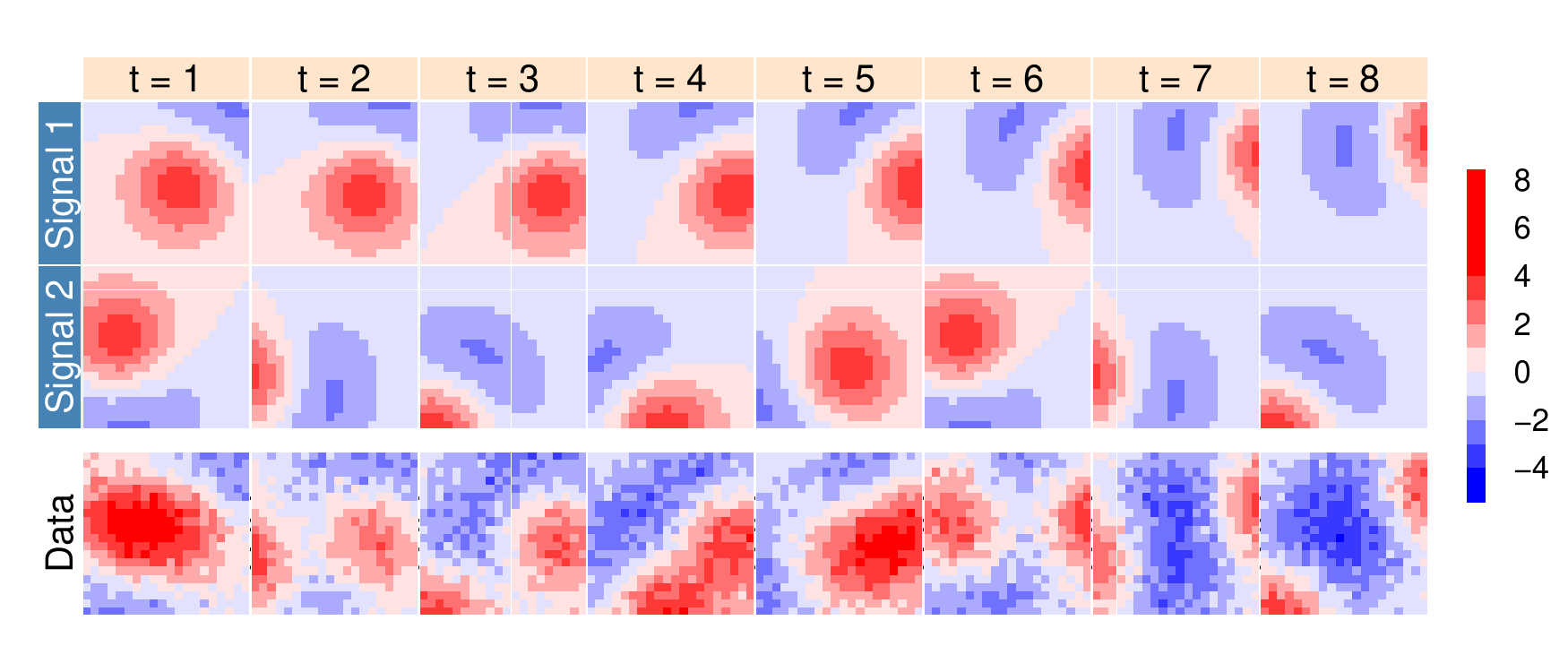}}
	\\
	\subfloat[][Mid-noise level: $\sigma^2_\varepsilon=4$]{
	\includegraphics[width=0.8\textwidth]{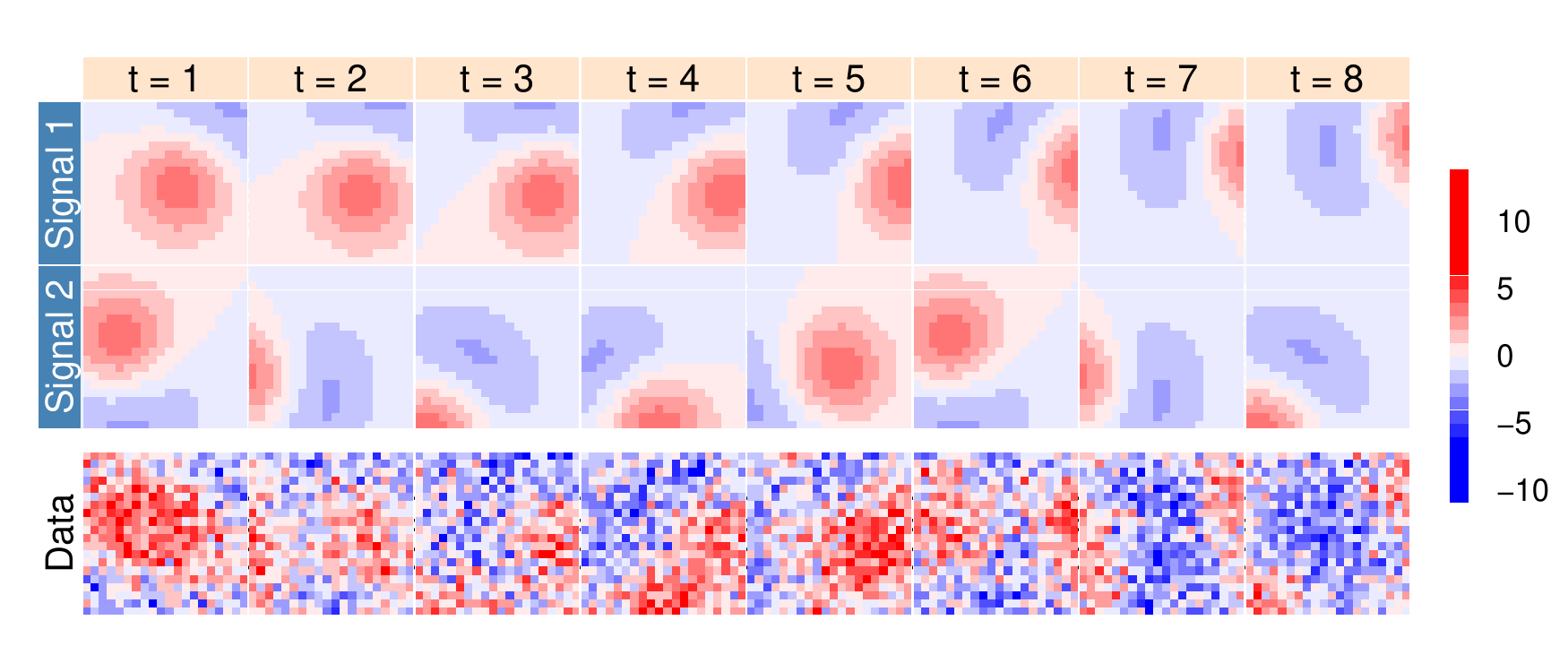}}
	\\
	\subfloat[][High-noise level: $\sigma^2_\varepsilon=16$]{
	\includegraphics[width=0.8\textwidth]{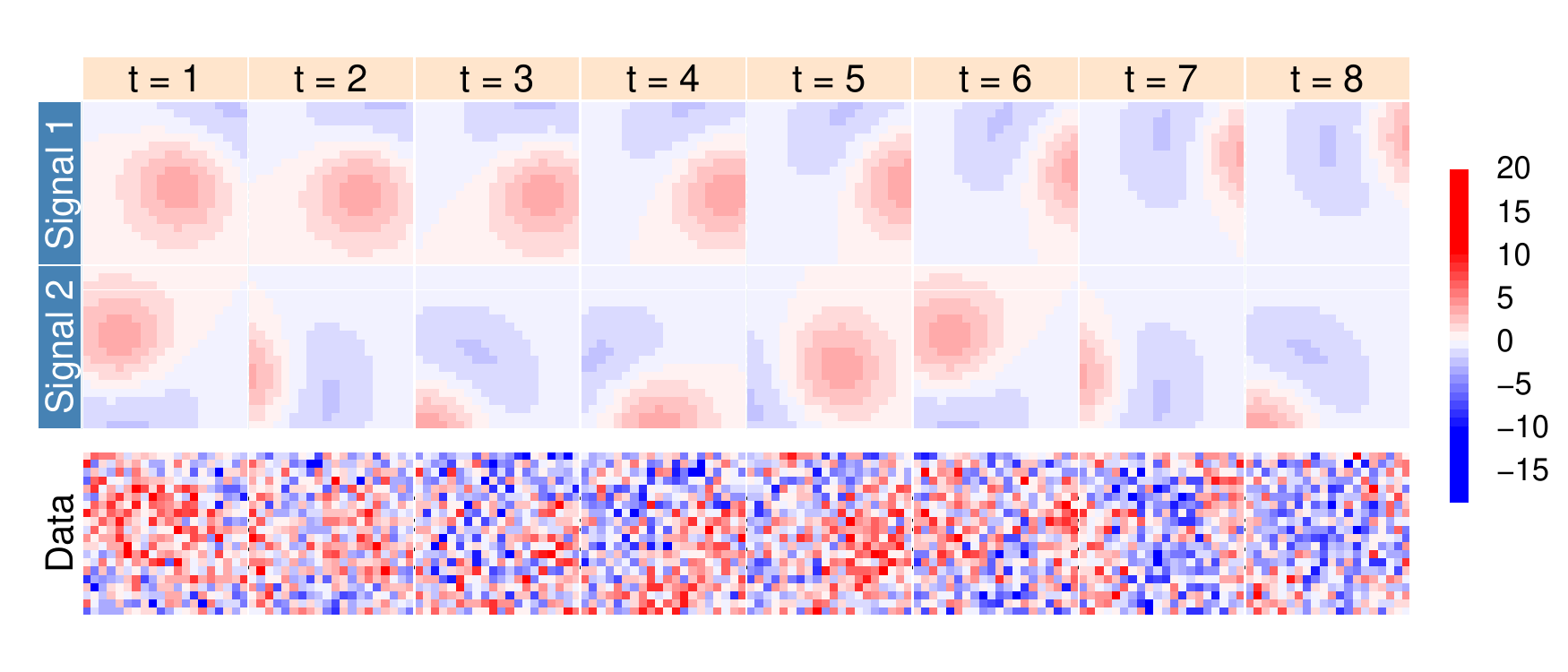}}
	\caption{Level plots of the two rotating energy sources and observed data at selected time in the low-, mid- and high-noise level settings.}
\label{fig:sim.twoball.data.all}
\end{figure}

\begin{figure}[hp]
	\centering
	\includegraphics[width=\textwidth]{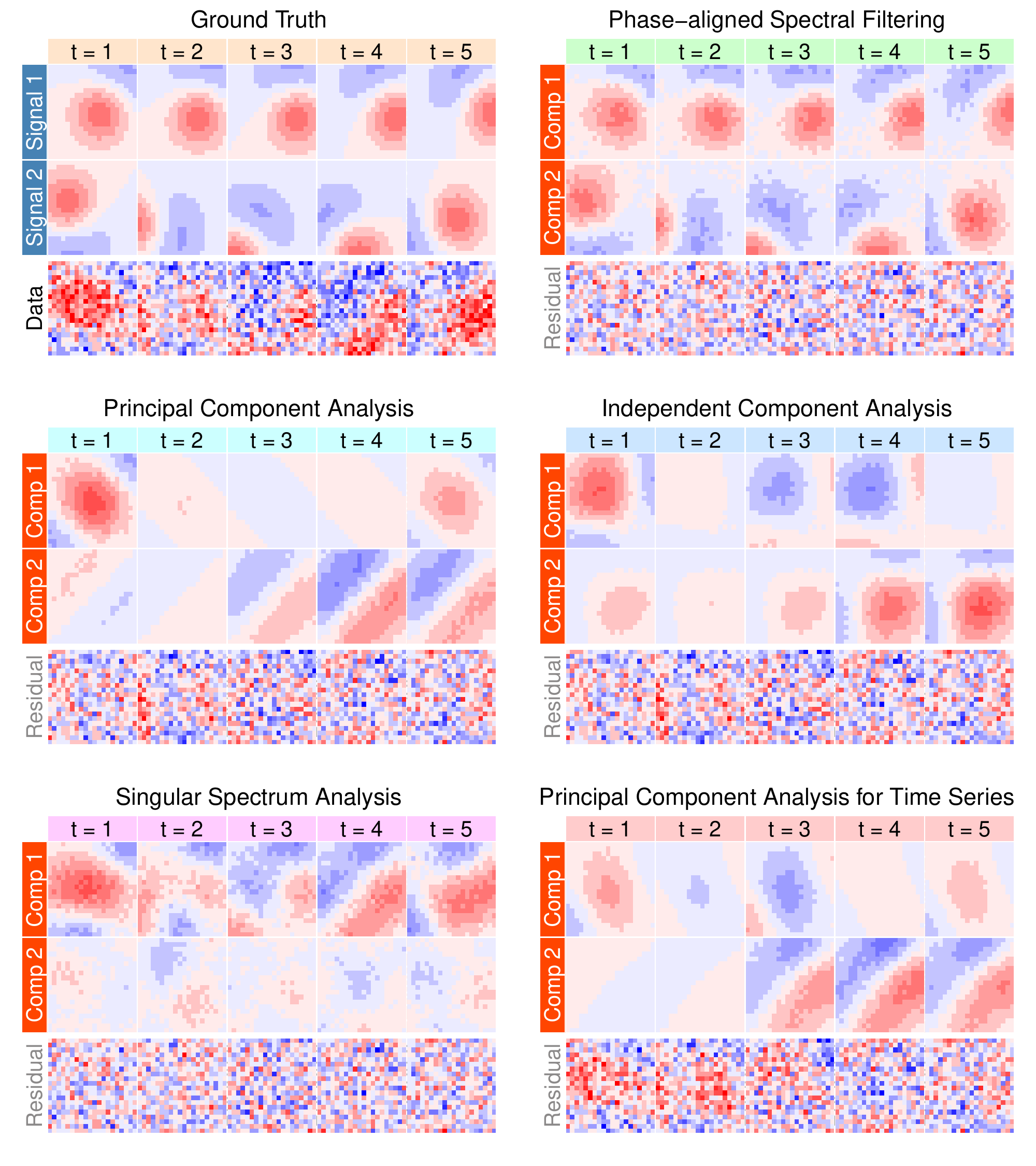}
	\caption{Decomposition results from different methods in the two-rotating-energy-source simulation in the mid-noise level setting with $\sigma_\varepsilon^2=4$.}
	\label{fig:sim.twoball.all.mid}
\end{figure}

\begin{figure}[hp]
	\centering
	\includegraphics[width=\textwidth]{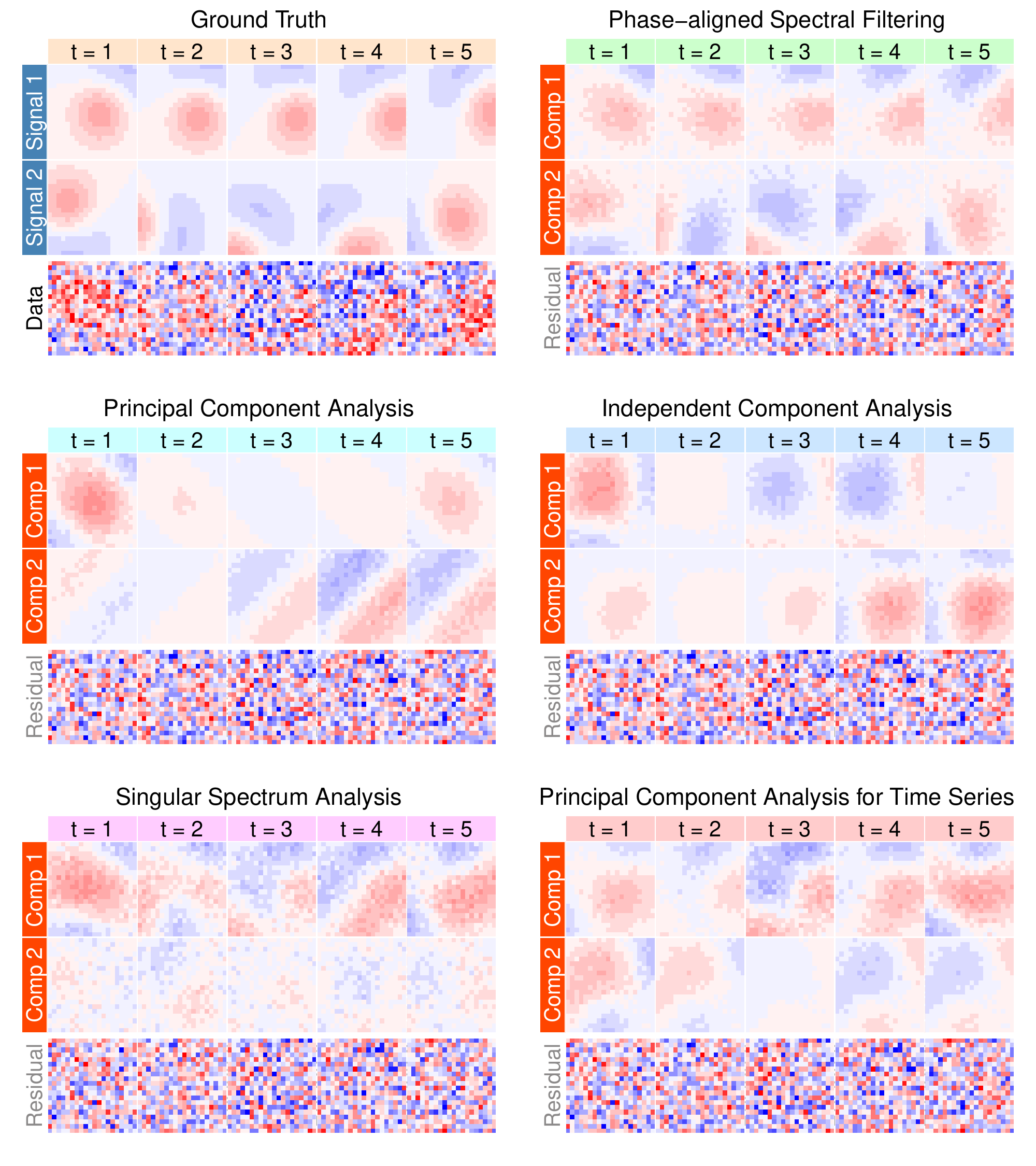}
	\caption{Decomposition results from different methods in the two-rotating-energy-source simulation in the high-noise level setting with $\sigma_\varepsilon^2=16$.}
	\label{fig:sim.twoball.all.high}
\end{figure}
	
\end{document}